\documentclass[twocolumn,showpacs,preprintnumbers,amsmath,amssymb]{revtex4}
\usepackage{tabularx,graphicx}\begin{document}
\newcommand{\beq}{\begin{equation}}
\newcommand{\eeq}{\end{equation}}
\newcommand{\beqn}{\begin{eqnarray}}
\newcommand{\eeqn}{\end{eqnarray}}
\newcommand{\bmath}{\begin{subequations}}
\newcommand{\emath}{\end{subequations}}
\title{
An index to quantify an individual's scientific research output that takes into account the effect of multiple coauthorship}
\author{J. E. Hirsch }
\address{Department of Physics, University of California, San Diego\\
La Jolla, CA 92093-0319}

\begin{abstract} 
I propose the index $\hbar$  (``hbar''), defined as the number of papers of an individual that have  citation count larger than or equal to the $\hbar$
of all coauthors of each  paper, 
as a useful index to characterize the scientific output of a researcher   that takes  into account   the effect of multiple coauthorship. The bar is higher for $\hbar$.
\end{abstract}
\maketitle 

\section{introduction and definition}
The $h$-index (number of papers of an individual with citation count $\geq h$) has gained considerable acceptance as a measure of individual research achievement that 
is advantageous compared to other bibliometric indicators such as total number
of citations or number of papers published in journals of high impact factor\cite{hreview,hreview2,fersht}. Various modifications of the $h$-index have been proposed to take into
account some of its perceived shortcomings\cite{h1,h1p,h1pp,h2,h2p,h2pp,h3,h4,h5,h6,h6p,h7,h7p,h8,schreiber,batista,egghe,chai}, however there is no general consensus so far that any other single number bibliometric indicator is clearly preferable to
the $h$-index.

The $h$-index can be defined as follows: {\it A scientist has index h if h of his/her papers belong
to his/her $h$-core. A paper belongs to the $h$-core of a
scientist if it has $\geq h$ citations}.
The $h$-core set is not necessarily
unique because there may be more than one paper with
citation count=$h$, but the $h$-index is.

In this paper I would like to define a new bibliometric indicator, $\hbar$, as follows:

 {\it A scientist has index $\hbar$ if $\hbar$ of his/her papers belong  to his/her  $\hbar$ core. A paper belongs to the $\hbar$ core of a scientist if it has 
$\geq$ $\hbar$ citations and    {\it    in addition}  belongs to the $\hbar$-core of each of the coauthors of the paper.}

The index $\hbar$ thus defined uniquely characterizes a scientist, as the $h-$index also does, and satisfies $\hbar \leq h$. For a scientist with only single-author papers,
$\hbar=h$, but $\hbar=h$ does not imply that the scientist only writes single-author papers, nor that only single-author papers form the $h$-core of this
scientist. 
Furthermore, the $\hbar$-index (unlike the $h$-index)  uniquely characterizes a  {\it  paper } as belonging or not belonging to the $\hbar$-core of its authors. 
(Instead, with the original $h$-index a multiple-author paper in general will belong to the $h$-core of some of its coauthors and not belong to the $h$-core of the 
remaining coauthors).   Thus, the scientific literature becomes divided into two non-overlapping sets, the $\hbar$-contributing papers and the
non-$\hbar$-contributing papers, and any given paper in the literature at a given time belongs to either set, never to both. As a function of time, a paper
could migrate back and forth between both sets, but the vast majority of papers will either never belong to the $\hbar$ set or remain in it once they enter it.
(For the nit-picking readers, ``belonging'' above should be
replaced by ``qualifying to belong'', since neither the $h$- nor the
$\hbar$-cores are unique\cite{burrell}).

The $\hbar$-index just defined is somewhat difficult to understand and in addition extremely difficult to calculate. Therefore, I will define
a ``non-self-consistent'' $\hbar$ by substituting $\hbar$ by $h$ in the last part of the definition, namely:

 {\it A scientist has index $\hbar$ if $\hbar$ of his/her papers belong  to his/her  $\hbar$ core. A paper belongs to the $\hbar$ core of a scientist if it has 
$\geq$ $\hbar$ citations and    {\it    in addition}  belongs to the $h$-core of each of the coauthors of the paper.}

In practice, the non-self-consistent $\hbar$ will be almost identical to the self-consistent one  (first definition),  occasionally slightly smaller. It loses some of the nice features
of the first  definition, in particular now a paper may belong to the $\hbar$-core of one of the coauthors and not to that of another one, although this will
 happen very infrequently. However with this modified definition
it becomes a practical bibliometric indicator that can be calculated by hand. Thus I will only deal with the non-selfconsistent $\hbar$ in the remainder
of this paper. I would like to propose $\hbar$ as a useful indicator to discriminate between scientists with different coauthorships patterns.

 \section{motivation}

Perhaps the most important  shortcoming of the $h$-index is that it does not take into 
account in any way the number of coauthors  of  each paper\cite{hindexorig,schreiber,batista,egghe,chai}. Thus, an author that publishes alone does not
get any extra credit compared to one that routinely publishes with a large number of coauthors, even though the time and effort invested per paper by the single author
or by each of the coauthors in a small collaboration is presumably larger than the corresponding one for a member of a larger collaboration. 
This can lead to serious distortions in comparing individuals with very different 
coauthorship patterns, and gives an incentive to authors to work in large groups when it is not necessarily desirable. For example, consider a group of several  researchers that decides to
put all of their names as authors in any paper they write, independently of how small or nonexistent the contribution of each of them was. The individual $h$-index of each of these researchers will 
be higher than it would have been in the absence of such decision. Of course such an extreme procedure would violate generally accepted standards of scientific integrity. Nevertheless, it is
sometimes a grey area whether or not a minor contributor should be included as author of a paper; with the $h$-index and other current bibiometric indicators there is no 
penalty to add authors to a paper and as a consequence there can be an incentive to do so, due to implicit or explicit quid pro quo expectations.

Thus, a bibliometric indicator that discourages honorary authorship and gives   extra credit to authors that publish alone or in small collaborations, and/or
 subtracts credit from coauthors in larger collaborations,
would be desirable. On the other hand, the indicator should not $discourage$ collaborations, that are essential
for the progress of science. This is the case for example in the modification of the $h$-index proposed by Schreiber\cite{schreiber} to take into account multiple coauthorship by 
counting the papers fractionally according to the inverse of the number of coauthors. Similarly, Batista et al\cite{batista} proposed to divide the $h$ index of an individual
 by the mean number of authors of the
papers in the $h$-core,   Egghe\cite{egghe} proposed to count either citations or ranking of papers in a fractional way to take into account the 
number of coauthors, and Chai et al\cite{chai} discussed a scheme to allocate partial credit to each coauthor of a paper.

Such modifications of the $h$-index unduly discourage collaborations in this author's opinion, in addition to not being necessarily accurate or fair. In particular, 
there are often large differences in the individual contributions of coauthors to a joint paper, so there is no good reason to divide the credit 
equally among coauthors\cite{dif1,dif2}. 
It will often be the case that in a paper with a large number of citations the senior author played a crucial role, and it would be
inappropriate to reduce his/her credit by an amount that depends on how many junior collaborators were involved in the project: 
the paper should fully contribute to  at the very least  that member of the collaboration that   conceived and led the project.
One could consider an indicator that gives  large credit to the first author of the paper, who often is the main contributor,
however different disciplines have different conventions
on ordering of author names, in some scientific subfields (e.g. high energy physics) being  almost always alphabetical.

One may be tempted  to consider an indicator that gives credit only to the leading author in a collaboration
(leading author meaning the leader of the collaboration, not the first author),  often (not always)
 a senior member of the collaboration.  However, it is 
not infrequently  the
case that there is more than one leading author  of a collaboration. It is also often the
case that very junior collaborators play a key role, in some cases they are listed as first authors and in other cases they are not.
In a paper that becomes so successful that it  eventually  becomes part of the $h$-core of even the most senior member of
the  collaboration it would certainly  be extremely  unfair to deny all credit to   junior collaborators that played a key role. Furthermore, such an indicator would completely discourage
junior scientists  from  collaborating with senior scientists, with a detrimental effect on the progress of science.

On the other hand, in comparing mid-career researchers with comparable citation numbers, it may be the case that one of them achieved most of his/her citations in papers 
where he/she was  the   leading author collaborating with junior coauthors (students and postdocs),
 and the other one achieved most of his/her citations as a minor contributor in collaborations with more senior authors.
For most evaluation purposes (eg awarding grant support or career advancement) the first researcher should be favored. This will  not happen  if   the
$h-$index is used as an indicator nor with the 
 proposed modifications of the $h$-index that exist in the literature.

A useful bibliometric indicator should (i) Reflect elements of reality that are useful for evaluation and meaningful    in a statistical sense (there are always exceptions to any criterion)
and ideally have predictive power\cite{hindex},
(ii) not lead to undesirable incentives that are detrimental to the progress of science, 
(iii) not be too sensitive to small variations in citation records that could be due to random events, 
and (iv) last but not least  be not  too difficult to obtain from existing databases.
I argue that $\hbar$ is a good candidate to meet these criteria and is superior to the $h$-index in that it takes into account the effect of
multiple coauthorship. $\hbar$ may be used alone or in combination with the $h$-index.

\section {calculation of $\hbar$}
To understand the meaning of $\hbar$, let us consider a hypothetical example. A researcher (A) has an $h$-index of 20, i.e. 20 papers with 20 or more citations each, papers number 1 through
20 in order of decreasing citations. Paper 20 has 21 citations and is single author, paper 19 has 25 citations and is authored by A and a junior collaborator who
himself has an $h$-index of 5. These papers are kept  in the $\hbar$ count.
Papers 18, 17 and 16     have 28, 30  and 35 citations and are coauthored with   senior scientists  B,  C and D respectively  who have h-indices 45, 55 and 40,
as well as  possibly other less senior (meaning of  smaller $h$-index) coauthors. Paper  
15 has 43 citations and is also coauthored with scientist D .

The  $\hbar$-index  will eliminate papers 18, 17 and 16 from the $h$-core of A because they are coauthored with senior authors whose $h$-index is higher
than the citation counts of those papers (and of course higher than the $h$-index of A) . It will keep paper 15 because it has 43 citations, thus contributing also to the $h$-core of the 
senior coauthor D (that has $h$-index 40).

Assume for the sake of simplicity that   all  the other papers in the $h$-core of scientist A (i.e. papers 1 through 14)  are  either single author or coauthored with only junior collaborators
(defined as collaborators with lower or equal $h$-index than that of A), so they  all contribute fully   to A's  $\hbar$ count. The original $h$-index of 20 is now reduced
by the 3 papers eliminated (18, 17, 16) to $\hbar_{first-iteration}=17$.

This is not the end yet, because there may  be other papers with citation counts  between 17 and 20 (not part of the original $h$-core). Assume paper 21 has 19 citations and is single author,
paper 22 has 19 citations and is coauthored with B, and paper 23 has 17 citations. Paper 21 should be added to the $\hbar$ count bringing it to 18,
paper 22 should not be added because it has fewer citations (19) than the $h$-index of coauthor B (45), and paper 23 with 17 citations is lower than
$\hbar=18$ so it does not contribute, as do not contribute the subsequent papers number 24, 25, 26, ... that have citation count lower or equal to 17 by definition.
Thus, for this scientist A, $h=20$, $\hbar=18$.

In summary: in the first iteration  some papers in the $h$-core are eliminated yielding $\hbar_{first-iteration}\leq h$.  If $\hbar_{first-iteration}<h$  a second and final
iteration is needed to possibly add   some papers not originally in the $h$-core to yield $\hbar$, with $\hbar_{first-iteration}\leq \hbar \leq h$.
All $\hbar$ papers of scientist A have citation count $\geq$  $\hbar$, and all of them contribute to the $h$-core of all the coauthors  of each paper except possibly
to the $h$-core of A him/herself (paper 21 in the above example). All the other papers of scientist A either have $\leq$ $\hbar$ citations or they 
have $> \hbar$ citations but don't contribute to the $h$-core of one of the coauthors of the paper.

The calculation of $\hbar$   is surprisingly simple to perform using e.g. the ISI database. Let us go through the steps. Go to ``General search'', enter the
name of the author. Sort  the papers by ``Times Cited'', accordingly paper numbers  $n_p=1,2,3...$ (leftmost numbers on the screen)   have decreasing number of citations $n_{cit}$ for 
increasing $n_p$, and find the paper number $h$
(i.e. the highest number paper $n_p$ that has  $n_{cit}\geq h$). Click on the title of the paper, leading to a page with the full reference and all the coauthors.
Click on each of the coauthors for which there is reason to suspect that his/her $h$-index   may be higher than $n_{cit}$. Find the $h$-index of each such
coauthor, $h_{c}$. Once (if) one is found with $h_c>n_{cit}$, this paper is eliminated from the $\hbar$ count and the remaining coauthors of this paper don't 
have to be checked. If none of the coauthors has $h_{c}>n_{cit}$, the paper remains in the $\hbar$-count.

This procedure is repeated for each paper going down in the paper number of the author (corresponding to increasingly cited papers). 
Once a paper is reached where its citations are clearly higher than that of any coauthor the procedure is stopped. For example, in physics we know that
the highest $h$-index is about $115$ at present, so papers with   $n_{cit}>115$ certainly  don't have to be checked for potential elimination.

When this procedure is completed one has identified a certain number of papers ($n_d$) that are deleted from the h-count. Now one goes back to paper number h,
and renames that paper as being number $n_p=h-n_d$, having $h$ or more citations. There will in general now be other papers with number $n_p>h-n_d$ that 
have number of citations   $n_{cit}\geq n_p$, which should be included in the $\hbar$ count $provided$ they are not eliminated because of  having highly
cited coauthors. Thus one continues in order of increasing $n_p$  to look for the additional papers to be included with   $n_{cit}\geq n_p$
until the crossing point is reached as in the usual $h$-index calculation. Each time a paper is eliminated due to highly cited coauthors, the numbering $n_p$ of
the subsequent papers (to be compared with $n_{cit}$)  is reduced by $1$.

Another path one can follow is to first click on the ``Analyze'' link to rank the records by the Author field. This will yield a list of all the coauthors of this author in order of decreasing
number of papers coauthored. By looking up the $h$-index of the coauthors at the top of this list one can simplify the procedure described above by knowing in advance 
which of the frequent coauthors are high $h$-index coauthors that should be watched in considering whether or not a paper is eliminated from the h-count
in computing $\hbar$.

\section {why is $\hbar$ useful?}

It is clear that computing the $\hbar$ index is considerably more time consuming than computing the $h$-index of a researcher. But it is clearly  straightforward and doable
in most cases in a few minutes. In particular, it is easy
for a researcher to compute his/her own $\hbar$ index, since presumably he/she has already a pretty good idea of what the $h$-index of most of his/her collaborators are.
It should also not be too hard to compute the $\hbar$-index of some of our competitors, that may have an $h$-index comparable to our own but somehow we have
the feeling that they are not as good as we are. Will the $\hbar$-index reflect this?

I argue that the $\hbar$-index further contributes to the   `democratization' of the evaluation and comparison of scientist's achievements,   just  as the $h$-index does compared for example to the impact
factor\cite{fersht}, and that it is likely to give a more accurate prediction of future achievement\cite{hindex}. Consider two junior researchers, one of which (A) had as thesis advisor a   senior, prolific
 and prominent scientist running a large research operation, and the other (B) had as thesis advisor a young 
 scientist with a small research operation. It is likely that researcher A has benefitted in getting many citations due to his/her collaboration with the
more prominent scientist and his/her large group of coworkers. In addition, the letters of recommendation written by the prominent advisor
of A are likely to have more weight with review committees than those written by  the more junior advisor of B.  If researcher A has a higher $h$-index than B, he/she may well be the better scientist.
But assume  A and B have comparable h-indices, then  it is likely that
scientist B has a higher $\hbar$-index, $and$ I argue that   it is likely that scientist B  has a more promising scientific future since the larger $\hbar$ index  accurately 
reflects his/her own higher individual abilities, not having benefitted as much as scientist A from external circumstances.
Quite generally, for two equally good papers authored by  a young researcher, the one with better known senior coauthors is likely to garner more citations simply
because of name recognition, thus it is reasonable that $\hbar$ counterbalances this effect.

It is certainly the case  that the $\hbar$-index   is particularly unkind to junior researchers. If the junior researcher has as coauthor a very senior researcher
that paper will not contribute to the  $\hbar$ count for many years even if it is an outstanding paper and even if the junior researcher's contribution to the
paper was seminal. But, if the paper is sufficiently good it will eventually contribute to the $\hbar$ count of both authors. Meanwhile, $\hbar$ should give junior researchers extra
incentive to devote efforts to their independent work, or to work with other even more junior or at most contemporary collaborators, which should have a positive effect in advancing their career
as well as in advancing new science. 
Nevertheless, in considering the bibliometric evidence for  junior researchers it is especially 
important to 
use the $\hbar$ index  $together $ with the $h$-index as well as other bibliometric as well as non-bibliometric criteria.
If one insists on a single bibliometric parameter perhaps an average of the $h$ and $\hbar$ indices would be  appropriate.

Note also that in collaborations where the authors have similar $h$-index, the collaboration will not result in a much reduced $\hbar$-index for any of the collaborators. For example, if the $h$-index
of coauthor B ($h_B$) is slightly higher than that of author A ($h_A$),  only the papers of A and B whose citation number fall in the narrow interval $[h_A, h_B]$ can be eliminated from the $h$-core of A in computing his/her $\hbar$. Thus, the $\hbar$-index does not penalize collaboration between scientists of similar seniority, unlike the other proposed modifications of the $h$-index to
take into account multiple coauthorship\cite{schreiber, egghe, batista,chai}.  Such collaborations are often very fruitful.

On the other hand, $\hbar$ clearly discourages honorary authorship for more senior/prominent researchers, and it also  discourages collaborations for the sole
purpose of increasing the coauthors' h- or $\hbar$-indices  even among collaborators with similar h-indices.  
Unlike the $h$-index, the $\hbar$-index of a researcher can decrease with time. 
This will happen if a coauthor's $h$ increases sufficiently so that a coauthored paper can
get eliminated from the $\hbar$ count. Even though for coauthors with similar h-indices 
their $\hbar$ indices are  not reduced, $\hbar$ could be reduced for one of them in the future if the coauthor's $h$-index increases substantially. 

The $\hbar$ index should be more accurate than the $h$-index in reflecting the individual contributions of a researcher where he/she played a leading role,
as well as those where he/she is likely to have played an important enough role to have   propelled the paper to the $h$-core of even its more senior coauthors.
As defined, the $\hbar$-index is probably almost useless for scientists at the stage of post-doc and of very limited use at the beginning assistant professor
stage. However I  believe that it
can start playing a significant role at the time of evaluation for promotion to tenure, where individual independent research contribution should be
evaluated, as well as further along in the scientist's career. Furthermore I argue that $\hbar$ should play a useful role together with the ordinary $h$-index and other criteria, for
decisions on allocation of research resources by granting agencies throughout a scientist's career except  perhaps  at the very beginning.

\section{examples} 

 \begin{table}
\caption{Publication data for a recently tenured physicist at UCSD: $h=17$, $\hbar=12$. $h_{max}$ denotes  the $h$-index 
of the highest-h collaborator of the paper (senior and single mean the physicist in question was senior author or single author of that paper). $n_p^\hbar$ counts the papers in the $\hbar-$core. The last column gives the publication year of the paper.}
\begin{tabular}{l || c | c  | c  | c  | c | c }
$n_p$ & $n_{cit}$ &$h_{max}$  & $\#$ of authors  &  $n_p^{\hbar}$&  $n_{cit}$  & paper year  \cr
\tableline
\tableline
1 &229 &  41 &3  &1 & 229 & 1996 \cr
\tableline
2 &186 &  41 &2  &2 & 186 & 1996 \cr
\tableline
3 &70 &  80 &3  &- & 70 & 2001  \cr
\tableline

4&67 &  41 &2  &3 & 67 & 1995  \cr
\tableline

5 &63 &senior   &2 & 4 & 63 & 1997 \cr
\tableline

6&43 &  41 &4  &5 & 63  & 1997\cr
\tableline

7 &39 &  41 &3  & - & 39 & 2004 \cr
\tableline

8 &28&  60 &3  & - &28 & 2000 \cr
\tableline

9  &26 &  41 &2 & - &26  & 2007\cr
\tableline

10 & 25 &  107 &3  & - &25 & 2006 \cr
\tableline

11 & 24 &  43 &5& - & 24 & 2008  \cr
\tableline

12 &23 &  senior & 4 & 6& 23 & 2008  \cr
\tableline

13 &22 &  41 & 2  & - & 22 & 1994  \cr
\tableline

14 &20 &  36 &3  & - & 20 & 2008  \cr
\tableline

15 &19 &  50 &6& - & 19 &2000 \cr
\tableline

16 &18 & single & 1  & 7& 18  & 2004\cr
\tableline

17 &17 &single&1 & 8 & 17 & 2002 \cr
\tableline
\tableline
18 &15 &  senior &2  &9& 15 & 2005  \cr
\tableline

19  &15 &  single&1  & 10 & 15 & 2005\cr
\tableline

20 &14 &  33 &7 & - & 14 &2006 \cr
\tableline

21 &13 &  single &1 & 11& 13 & 2005 \cr
\tableline

22 &12 &  senior  &2& 12 & 12 & 2005 \cr
\tableline
\tableline
23 &11 & 28&3& -& 11 & 2006 \cr
\tableline
24 &10 & 41 & 2 & -& 10 & 1998 \cr
\tableline
25&9 & single&1& 13& 9 & 1998 \cr
\tableline

\end{tabular}
\end{table}

Let us now consider some examples. 
 As a first  illustrative example I give in table I the publication record (25 highest cited papers)  of a recently tenured physicist at UCSD. His scientific age is $n=15$, total number of papers $51$, total citations $1074$,
$h=17$ and $\hbar=12$. $h_{max}$ in each row gives the highest $h$-index among the coauthors of that paper, ``senior'' and ``single'' mean that the physicist in question is either the senior member of the collaboration or the single author,
hence equivalent entries would read $h_{max}=17$. It can be seen that the 12 papers contributing to $\hbar$ are a healthy mix of very highly cited papers with senior coauthors
($n_p=1, 2, 4$ and $6$), and papers where this scientist is senior or sole member of the collaboration ($n_p=5, 12, 16, 17, 18, 19, 21, 22$). Highly cited paper number 3 (70 citations)  is eliminated
from the $\hbar$-count due to the very high $h$ of a collaborator, $h=80$, as are several others ($n_p= 7, 8, 9, 10, 11, 13, 14, 15$) . 
Several of these  coauthored papers that are presently
 part of the $h-$core but not of the $\hbar$-core  will eventually become part of the $\hbar$-core once/if their number of citations exceeds $h_{max}$ for that row
 (which will of course also increase  with time in general). Four papers that were not in the $h$-core ($n_p=18, 19, 21, 22$)  get added to the $\hbar$-core, due to the fact that only $8$ out of $17$ papers in the $h$-core survived the
 extra requirement needed to belong to the $\hbar$-core. These are all fairly recent senior- or single-author papers, indicating that the researcher is becoming increasingly independent.

 It would be unfair and counterproductive in this author's opinion to make this researcher ``pay'' in his bibiliometric index because of having collaborated with his graduate students and postdocs in his
 papers $n_p=5, 12, 18, 22$ instead of being sole author, as would be the result of applying any of the fractional credit schemes proposed in the literature to account for
 multiple coauthorship\cite{schreiber,batista,egghe,chai}.
 
It is possible in this example that the self-consistent $\hbar$ is slightly higher than the non-self-consistent one: the easiest way for this to happen would be if
$h_{max}=41$ in paper $n_p=7$ would correspond to an $\hbar$ for that scientist of $39$ or lower,   in which case that paper would belong to the self-consistent $\hbar$ core
of all the coauthors of the paper including the scientist under consideration.

D.J. Scalapino is a highly productive leading senior researcher in condensed matter theory ($h=81$)  that has worked with a large number of
collaborators. As a second set of examples,
Table II shows the $h$ and $\hbar$ indices for Scalapino and 25 of his collaborators (identified by their initials), with the last column giving the scientific age of the scientist (time since first published paper). Scalapino himself has $\hbar=h$, being the most senior researcher in  all his highly cited coauthored papers. The $h$ indices of 
Scalapino's collaborators in table I range from 59 down to 7, the $\hbar$ indices from 58 to 1, and $\Delta h=h-\hbar$ for a given 
researcher ranges from 10 to 1. It should be pointed out that the reduction from $h$ to $\hbar$ in the various cases occurred not only because of collaboration with
Scalapino but also with several other senior researchers in condensed matter theory.
Table II is arranged in order of decreasing $h$ index.

The following features are interesting to note: 

(i) There is not a very clear correlation between $\Delta h$ and $h$. For example, SRW with $h=47$ has $\Delta h=10$, EJ with $h=14$ has
$\Delta h=4$. However, it is true that very junior researchers have a very large $\Delta h$ compared to their $h$.

(ii) There is also not a clear correlation between $\Delta h$ and $n$, scientific age. ED with $n=26$ has $\Delta h=1$, 
RLS with $n=42$ has $\Delta h=5$. Thus it is certainly not generally true that very  senior researchers will always have 
converging $h$ and $\hbar$.

 It can be seen from the $\hbar$ column that in several cases the $h$ and $\hbar$ ranking orders are reversed, 
in particular:

(iii) SRW has $h=47$, larger than RLS's $h=43$, yet RLS's $\hbar=38$ is larger than SRW's $\hbar=37$.

(iv) DP's $h=35$ is larger than that   RTS's h=34, MJ's h=34 and WH's h=34, yet these researcher's $\hbar$
of 29, 33 and 30 are all larger than DP's $\hbar=27$.

(v) MJM's h=21 is smaller than NEB's h=23, yet MJM's $\hbar=19$ is larger than NEB's $\hbar=17$.

(vi) EJ's h=18 is smaller than NB's h=20, PM's h=20 and TD's h=21, yet  
EJ's $\hbar=14$  is larger than NB's $\hbar=12$, PM's $\hbar=13$ and TD's $\hbar=11$

\begin{table}
\caption{$h$ and $\hbar$ indices for D.J. Scalapino (condensed matter theorist, $h=81$) and $25$ of his collaborators. $\Delta h \equiv h-\hbar$, and $n$ is the scientific age $\equiv$ number
of years since first published paper.}
\begin{tabular}{l || c | c  | c  | c  }
Researcher  & $h$ & $\hbar$  &  $\Delta h$& n(years)   \cr
\tableline
\tableline
DJS &   81 & 81  & 0& 45  \cr
\tableline
ED & 59 & 58 &  1  & 26  \cr
\tableline
SRW &47 & 37 &  10  & 23    \cr
\tableline
AM& 43 & 35 &  8  & 25  \cr
\tableline
RLS & 43 & 38 &  5  & 42  \cr
\tableline
DP & 35 & 27 &  8  & 23 \cr
\tableline
RTS& 34 & 29 &  5  & 24 \cr
\tableline
MJ & 34 & 33 &  1  & 23  \cr
\tableline
WH& 34 & 30 &  4  & 38  \cr
\tableline
AVB & 30 & 26 &  4  & 25  \cr
\tableline
PJH& 27 & 25 &  2  & 23  \cr
\tableline
AWS & 27 & 23 &  4  & 18  \cr
\tableline
JEG & 25 & 23 &  2  & 38  \cr
\tableline
JKF & 24 & 21 &  3  & 21  \cr
\tableline
NEB & 23 & 17 &  6  & 24  \cr
\tableline
TPD & 23 & 20&  3  & 18  \cr
\tableline
RMN & 21 & 16 &  5  & 18  \cr
\tableline
TD& 21 & 11 &  10  & 17 \cr
\tableline
MJM& 21 & 19 &  2  & 23 \cr
\tableline
PM & 20& 13& 7 & 19  \cr
\tableline
NB & 20 & 12 &  8  & 20  \cr
\tableline
EJ& 18 & 14 &  4  & 16  \cr
\tableline
FFA & 16& 10 &  6  & 19  \cr
\tableline
LC & 14 & 8 &  6  & 12  \cr
\tableline
TAM & 11 & 5 &  6  & 7  \cr
\tableline
SG & 7 & 1 &  6  & 7  \cr
\tableline

\end{tabular}
\end{table}

Furthermore, in several cases where the h-indices are identical for two researchers the $\hbar$ indices are quite different, for example:

(vii) AM and RLS have both h=43, but RLS's $\hbar=38$, AM's $\hbar=35$.

(viii) MJ, RLS and WH all have $h=34$, yet their $\hbar$'s are 33, 30 and 29 respectively.

(ix) PJH and AWS have both $h=27$, yet their $\hbar$'s are 25 and 23 respectively.

(x) TPD and NEB have both $h=23$, yet their $\hbar$'s are 20 and 17 respectively.

(xi) MJM, RMN and TD have all $h=21$, yet their $\hbar$'s are 19, 16 and 11 respectively.

Note that MJ, MJM and EJ have particularly small values of $\Delta h$ compared to their similar-age peers. 
When reordering  the entries in Table II according to decreasing $\hbar$ rather than decreasing $h$, 
they  move up 3 ranks. I suggest that their higher $\hbar$ values reflects higher individual accomplishment and promise of
future accomplishment than is reflected in their $h$-index, i.e. compared to peers with similar $h$-index and larger $\Delta h$.
The future will tell.

As a final example I discuss the case of high energy experimental physicists. For these scientists the $h$-index is 
not very meaningful  because they
usually work in collaborations with hundreds of physicists, and everybody's name is listed in the author's list. The $h$-index of mid-career and senior successful
high energy experimentalists is typically in the 40's and higher. However, it is reasonable to expect that their $\hbar$ index will be much lower. To test this expectation
I attempted to calculate the $\hbar$ index of a senior high energy experimentalist at my institution, that has $n=34$ (first paper published in 1975),
334 publications, and h=44. Going down from $n_p=44$ progressively to lower $n_p$ values the citation count increases very slowly and 
papers are eliminated because of coauthors with higher $h$, eg. G. Coignet (h=49), D. MacFarlane (h=54), J.D. Burger (h=64). 
Paper number 18 in the publication list has 63 citations (in going from paper 44 to paper 17, citation count increases only from 44 to 64!), so 
paper 18 gets eliminated
from the $\hbar$-count because Burger is a coauthor. Therefore, the first iteration yields $\hbar=17$ $or$  $lower$.
Assuming no paper below $n_p=17$ gets eliminated (I did not check that fully), it is now necessary to consider the papers with
$n_p=45$ and higher to see if there are papers with $n_{cit}>17$ that don't get eliminated by having high $h$ coauthors.
One finds that papers $45$ to $166$ have $n_{cit}$ ranging from 43 down to $18$, paper $167$ has $n_{cit}=17$, and all these papers are
eliminated by having coauthors with $h$ higher than $n_{cit}$, not surprisingly (since almost every paper has hundreds of coauthors).
The conclusion is then that for this researcher $h=44$ and $\hbar=17$ (or even lower in case there are higher $h$ coauthors in 
papers 1 to 17 that I missed), i.e. $\Delta h=27$. It is reasonable to expect that such a large reduction from $h$ to $\hbar$ will be the norm for
high energy experimental physicists, as well as in other fields where collaborations typically involve a very large number of coauthors.

\section{closing arguments}

$\hbar$ gives full credit to the senior coauthor of a paper (senior meaning the coauthor with highest $h$-index), where by `full credit'  I mean the
same credit that $h$ would give. $\hbar$  may or may not give any credit to a junior coauthor of a paper who does get credit for that paper in
his/her  $h$-index.
Thus, $\hbar$ will give deserved extra credit  to those young and mid-career scientists that lead vigorous independent research programs compared to those that don't, encourage them to
take on younger students and postdocs without any penalty, and discourage them from 
instead spending a lot of their effort
in collaborations in research projects led by   more senior scientists, as well as discourage them from including the name of senior/prominent scientists in
their papers as coauthors (as opposed to e.g. in an acknowledgement for a minor contribution)  for purely political reasons. The author believes that these incentives are fair and   beneficial to the progress of science. It is likely that scientists with higher $\hbar$ than colleagues with the same $h$ will make better use
of research funds allocated to them by granting agencies. Making such authors ``pay'' in their bibliometric indices\cite{schreiber,batista,egghe,chai} for working with students and postdocs is neither fair nor beneficial to anybody 
nor does it yield any useful information about these scientists.

It may be reasonably argued that 
for junior scientists, papers with senior coauthors should be taken into account at least fractionally while they are part of the $h$-core but not
of the $\hbar$-core rather than not counted at all. That is easily achieved by using 
as indicator a  weighted average
(for example  the arithmetic average with equal weights)
of $h$ and $\hbar$ as defined here. The weights in such an average may
even be taken to be time-dependent, with the relative $\hbar$-weight starting from
zero and increasing with time. It may  be argued that whether or not a 
paper in the $h$-core makes it to the $\hbar$-core should   depend on the
publication date of the paper relative to the present. However such algorithms would
become very complicated.

To the extent that grant awarding agencies and faculty recruitment and   advancement  committees consider the information provided by the $h$-index in awarding grant support and evaluating 
job candidates and career advancement, I believe it is imperative that they consider
  the information provided by $\hbar$ also. A young scientist whose $h$ doesn't quite measure up to that of his/her peers should definitely get
  a second look if his/her $\Delta h$ is particularly small.  It would be both unfair and not conducive to optimal results (i.e. advancement of the best science) to favor a candidate that achieved a high 
  $h$ index predominantly by joining collaborations with  prominent senior colleagues over another one that achieved perhaps a somewhat smaller $h$ but a
  substantially higher $\hbar$ through mostly leading his/her own
  independent  research effort with junior colleagues. In addition, I suggest that the papers in the $\hbar$-cores should be
  particularly scrutinized in more detailed evaluation and comparison between scientists.

$\hbar$ is very harsh  on the very young scientist, and  as   mentioned earlier
 it should only be used if at all in such cases in combination with the $h$-index and other indicators, as well as with other non-bibliometric criteria.
One may fear that $\hbar$ will unduly discourage   young scientists from collaborating with  senior and/or  prominent scientists. However, there are plenty of other incentives
for young scientists to collaborate with senior/prominent scientists, namely the availability of resources, the benefits of learning from 
top scientists, and the letters of recommendation to be obtained from these influential members of the scientific establishment, to name
just a few. 
In the light of $\hbar$,  the young scientist collaborating with senior scientists should regard his/her  effort as a long term investment,
that may pay off (in increasing his/her $\hbar$) eventually, in addition to providing the shorter term benefits just mentioned.

There will undoubtedly be cases where $\hbar$ will do injustice. For example, a young theorist 
that interacts closely with a senior experimentalist and makes suggestions for new data taking may write theoretical papers explaining the experimental data where
the experimentalist is a coauthor, and the paper will not contribute to the theorist's $\hbar$-count for many years;  another similar theorist interacting less closely with the experimentalist may just use the available experimental data 
and thank the experimentalist in an acknowledgement, thus having the paper count much earlier to his/her $\hbar$-index.
Such situations are obviously beyond the grasp of any necessarily coarse-grained bibiliometric indicator, and highlight the need for evaluators
to consider individual circumstances in each case in addition to the indicators.

$\hbar$ gives full credit to all members of a collaboration once the paper garners enough citations to contribute to 
everybody's $h$-core. It may be argued that even in that case the contribution of some members of the collaboration might have been
so insignificant that the credit is undeserved. This may be true in some cases, but more often than not in very successful papers all
participants are likely to have played important roles in making the paper successful, and if not, well - so be it.

$\hbar$ is not very friendly to high energy experimentalists  nor presumably to researchers in other fields where collaboration among a very large number of
scientists is the norm. I believe this means that a 
low $\hbar$ value for a high energy experimentalist (low compared to a non-high-energy experimentalist) should not be interpreted as
shedding negative light on the scientist,  and a 
high $h$ value for a high energy experimentalist 
(high compared to a non-high-energy experimentalist) should not be interpreted as
shedding positive light on the scientist. In a nutshell, neither  $h$ nor  $\hbar$ indices are  very useful for high energy experimentalists except perhaps for
comparison with other high energy experimentalists, and the same should be true for other regular large-group collaborators.

It should   be interesting  to explore the time evolution of $h$ and $\hbar$ for individual scientists. At the beginning of a scientist's career there 
 will usually be a large difference ($\hbar$ much smaller than $h$), and as the career advances 
 the ``hbar-gap'' should gradually close  ($\Delta h \rightarrow 0$) as 
 $\hbar$ and $h$ indices  converge, the more so the
more independent and successful the scientist is. As an extreme  example, for Ed Witten, who has the highest $h$-index among physicists, it is clear by definition
that $\hbar=h$ (unless he collaborated with e.g. a biologist with much higher $h$-index). It is likely that almost all leading senior
scientists have $\Delta h=0$ or very small. Moreover, in comparing scientists across disciplines\cite{compare}  that have different citation patterns and
different typical values of $h$, scientists with $\Delta h\sim 0$ will typically be   the leaders in their discipline.

Note that the $\hbar$-index will ``weed out'' from the individual $h$-index predominantly those papers that have citation count close to $h$, and will always keep those
papers with very large number of citations independent of their authorship. Therefore it will give relatively higher weight to very highly cited papers relative to 
lower cited papers, both of which contribute equally to the $h$-index. This is similar to what is aimed at in some of the proposed modifications to the $h$-index,
e.g.  Egghe's  g-index\cite{h1}, that weighs highly cited papers more.

In fact, the concept of the $\hbar$ index can also be applied in an identical way to any of the other indices that have been proposed as alternatives to
the $h-$index. For example, Egghe's g-index   defines the g-core as the highest number of papers g that received $on$ $average$ g or more citations.
The  ``gbar'' index would add  the requirement that each paper should belong to the ``gbar'' core of all its coauthors (or to the g-core for the
non-self-consistent version).

As mentioned earlier, the $\hbar$ index can decrease with time. A young scientist collaborating with similar-aged peers may see
his/her $\hbar$-index gradually  $decrease$ in later years if his/her former collaborators vastly outperform him/her in scientific achievement in later years.
Unlike with $h$, ``we must keep running to stay in the same spot''.

The calculation of (the non-self-consistent) $\hbar$ using information provided in the ISI and Scopus databases is feasible with moderate effort (typically a few minutes compared to a few seconds
for the calculation of $h$). It would be facilitated if ISI and Scopus would provide the $h$-indices with fewer needed clicks, e.g., if in clicking on the paper title the list of authors appeared together with their
respective $h$-indices. That should not be very computer-time consuming. Currently in ISI one needs two further clicks to reach the $h$-index of each coauthor (clicking on the coauthor's name, and then on the ``Create Citation Report'' link). 
The latter can be quite time-consuming because it calculates many other things in addition to the $h$-index, and in fact it is often much faster to calculate the $h$-index of each coauthor by hand, as 
originally described\cite{hindexorig}. Alternatively, it would  greatly facilitate the calculation of $\hbar$
 if ISI would provide the $h$-index of the coauthors when  using the ``Analyze'' link and
ranking the records by the Author field.  However it would probably  be dangerous to leave the entire calculation of $\hbar$ to the computer due to
disambiguation issues: a computer-calculated $\hbar$ can be trusted as a lower bound, however one would have to
check by hand for the possibility that  the $h$ of a coauthor might have been inflated by including different scientists with the same name,
thus  depressing $\hbar$ more than warranted.

The self-consistent $\hbar$ is an appealing theoretical construct but has no practical value since it appears 
extremely difficult to compute, involving an iterative process including an ever-growing number of scientists. It is however likely to be identical or very close to the non-self-consistent $\hbar$
considered here in almost all cases, always an upper bound. One could also define an index that counts only the papers in the $h$-core of each author of the paper including the author under consideration, however such an index would give unreasonably low values in many cases (e.g.
$8$ rather than $12$  in the example of table I).

In the paper where I introduced the $h$-index\cite{hindexorig}, I suggested that in physics reasonable values of $h$ 
(with large error bars) 
might be $h\sim12$ for advancement to tenure (associate professor),
$h\sim18$ for advancement to full professor, $h\sim 15-20$ for fellowship in the American Physical Society and $h\sim 45$ for membership in the 
US National Academy of Sciences (NAS). In hindsight, the estimate for fellowship in the American Physical Society was much  too  low,
$h\sim 20-25$ being more typical values\cite{hindex}. One may wonder about similar estimates for $\hbar$. For membership in NAS it should not be very different, 
since it is usually late in a scientist's career. If the $\hbar$ of a scientist proposed for membership of the NAS is much smaller than
his/her $h$, this should raise red flags about his/her independence. For fellowship in the APS $\hbar$ should perhaps be $\sim 17-22$ and for advancement to full professor 
perhaps $\hbar\sim 15$ or $16$.  For tenure, an $\hbar$ as low as $7$ or $8$ may be sufficient 
especially if $h$ is substantially higher.

One of the attractive features of the $h$-index is that equally good papers (i.e. papers acquiring citations at the same rate) will give credit to a young
author much faster than they do to an older author, that has a higher $h$ bar to overcome. That attractive feature is preserved in $\hbar$ if there are
no senior coauthors and lost if there are.  But, on the other hand it may be argued that a young person has the luxury of many more years ahead
in his/her career, and in a collaboration involving junior and senior coauthors
a good paper may  contribute to the $\hbar$-count of the young coauthor when it is already too late
for the senior coauthor to benefit from it in either $h$ or $\hbar$ (i.e. after retirement or worse). 

An interesting open question is what is the relative size of $\hbar$-contributing $(\hbar$-core) versus $\hbar$-non-contributing papers in a scientific field or subfield
over a given time period.
Are they similar or very different between fields where the citation rate may vary greatly? 
It should not be too difficult to obtain estimates for these quantities by appropriate statistical sampling.
It is tempting to conclude that in a field where the $\hbar$-core is
say $5\%$ (i.e. $95\%$ of papers are being essentially ignored)
resources are being wasted compared to another field where the $\hbar$-core is say $30\%$. 
If such large differences exist, do they  correlate with perceptions of fields being stagnant versus rapidly advancing?
 Such quantitative  information could yield interesting insight
for decisions on funding allocations among fields by granting agencies, institutions and even countries.

Young scientists with an independent bent will probably like $\hbar$.
Those that continue coauthoring with their  thesis or postdoctoral advisors for many  years, and those that join high profile collaborations wherever they can find them, will probably 
not be thrilled by it. Established scientists that benefit from collaboration with young coauthors
even when their (the established scientist's) participation is minor will probably dislike it, since to the extent it is used as an evaluation tool it will make junior authors
more reluctant to include senior coauthors in their papers. 
No bibliometric indicator will (nor should) make everybody
happy.
To the extent that $\hbar$ gets acceptance, smart students will have a larger incentive to join a research group led by a young faculty member at the height of his/her creative period but with not yet
a  very high $h$, versus joining the group
of a prominent senior professor that may already be on the decline; the higher influence of the letter of recommendation of the senior professor will be 
tempered by the larger toll in $\hbar$ excised   from  the members of such group. These  incentives would appear to be beneficial to the vitality and
innovation quality  of the scientific enterprise. In the end, the effect of  $h$ and $\hbar$ and other bibliometric indicators   on the scientific enterprise will depend on how they are used,
and hopefully the scientific community and the scientific administration community will converge to the right uses  of the right indices to 
provide optimal incentives for the advancement of science.

The bar is higher for  $\hbar$. For a paper that counts for the $h$ of a scientist to also count for his/her  $\hbar$, either the scientist
has to have been the  sole author or the senior author of  the paper, or else the scientist together with his/her  
other junior and senior coauthors should have delivered enough bang for the buck to make it a significant paper for each and every one of the coauthors involved.  
A vanishing  hbar-gap in the advanced  stages of a scientist's career is a hallmark of scientific leaders.

\acknowledgements 
The author is grateful to Dr. Prabhat  Koner for stimulating correspondence.


\begin{references}
\bibitem{hreview} 
Alonso S., Cabrerizo F.J., Herrera-Viedma E., Herrera F. (2009) h-Index: A review focused in its variants, computation and standardization for different scientific fields.  Jour. of Informetrics 3: 273-289  and references therein.
\bibitem{hreview2} Bornmann L., Daniel HD (2007) What do we know about the h index? Jour. of the Am. Soc. for Information Science and Technology  58: 1381-1385.
and references therein. 
\bibitem{fersht} Fersht A. (2009) The most influential journals: Impact Factor and Eigenfactor. PNAS  106: 6883-6884.
\bibitem{h1} Egghe L. (2006) Theory and practise of the g-index.  Scientometrics 69: 131-152.
\bibitem{h1p}  Van Raan A.F.J. (2006) Comparison of the Hirsch-index with standard bibliometric indicators and with peer judgment for 147 chemistry research groups. Scientometrics 67: 491-502.
\bibitem{h1pp}  Komulski M. (2006) A new Hirsch-type index saves time and works equally well as the original h-index.  ISSI newsletter 2: 4-6
\bibitem{h2} Jin B.H., Liang L.M., Rousseau R., Egghe L. (2007) The R- and AR-indices: Complementing the h-index. Chinese Science Bulletin 52: 855-863.
\bibitem{h2p}  Jin B. (2007) The AR-index: complementing the h-index.   ISSI Newsletter3: 6-6.
\bibitem{h2pp} Sidiropoulos A., Katsaros D.,  Manolopoulos Y.  Generalized Hirsch h-index for disclosing latent facts in citation networks.
Scientometrics 72: 253-280.  
\bibitem{h3} Bornmann L., Mutz R.,  Daniel, H.D. (2008)  Are there better indices for evaluation purposes than the h index? a comparison of nine different variants of the h index using data from biomedicine. Jour. of the Am. Soc. for Information Science and Technology 59: 830-837.
\bibitem{h4} Van Eck N.J. , Waltman L. (2008) Generalizing the h- and g-indices.  Jour. of Informetrics 2: 263-271.
\bibitem{h5} Antonakis J., Lalive R. (2008) Quantifying scholarly impact: IQp versus the Hirsch h. 
Jour. of the Am. Soc. for Information Science and Technology 59: 956-969.  
\bibitem{h6} Ruane F., R. Tol R. (2008) Rational (successive) h-indices: An application to economics in the Republic of Ireland.  Scientometrics 75: 395-405.
\bibitem{h6p} Rousseau R., Ye F. (2008) A proposal for a dynamic h-type index. Jour. of the Am. Soc. for Information Science and Technology 59: 1853-1855. 
\bibitem{h7} Anderson T., Hankin K., Killworth P.  (2008) Beyond the durfee square: Enhancing the h-index to score total publication output. Scientometrics 76: 577-588.
\bibitem{h7p} Egghe L., Rousseau R. (2008) An h-index weighted by citation impact.  Information Processiong and Management 44: 770-780.  
\bibitem{h8} Guns  R. , Rousseau  R. (2009) Real and rational variants of the h-index and the g-index. Jour. of Informetrics 3:  64-71.
\bibitem{schreiber}  Schreiber  M. (2009) A case study of the modified Hirsch index $h_m$ accounting for multiple coauthors. 
Journal of the American Society for Information Science and Technology 60: 1274-1282.
\bibitem{batista} Batista P.D., Campiteli, M.G., Kinouchi O., Martinez A.S. (2006) 
Is it possible to compare researchers with different scientific interests? Scientometrics 68:179-189.
\bibitem{egghe} Egghe L.  (2008) Mathematical theory of the h- and g-index in case of fractional counting of authorship. Journal of the American Society for Information
Science and Technology 59: 1608-1616.
\bibitem{chai} Chai J.C., Hua P.H., Rousseau R., Wan J.K. (2008) Real and rational variants of the h-index and the g-index. 
Proc. of WIS 2008, Berlin, H. Kretschmer $\&$ F. Havemann, eds., 64:71.
\bibitem{burrell} Burrell Q.L.  (2007) On the h-index, the size of the Hirsch core and Jin's A-index. Jour. of Informetrics 1:
170-177.
\bibitem{hindexorig} Hirsch J.E. (2005) An index to quantify an individualÕs scientific research output. PNAS 102: 16569-16572.
\bibitem{dif1} Sekercioglu C. H. (2008) Quantifying Coauthor Contributions. Science 322: 371-371.
\bibitem{dif2}  Zhang C.T.  (2009) A proposal for calculating weighted citations based on author rank. EMBO reports 10: 416-417.  
\bibitem{hindex} Hirsch J.E. (2007) Does the h index have predictive power? PNAS 104: 19193-19198.
\bibitem{compare}      Radicchi F., Fortunato S., Castellano C.  (2008)
Universality of citation distributions: toward an objective measure of scientific impact.  PNAS  105: 17268-17272.   
\end{references}
\end{document}